\documentclass[letterpaper]{article} 
\usepackage{aaai24}  
\usepackage{times}  
\usepackage{helvet}  
\usepackage{courier}  
\usepackage[hyphens]{url}  
\usepackage{graphicx} 
\usepackage{natbib}  
\usepackage{caption} 
\usepackage{algorithm}
\usepackage{algorithmic}

\usepackage{newfloat}
\usepackage{listings}
\DeclareCaptionStyle{ruled}{labelfont=normalfont,labelsep=colon,strut=off} 
\floatstyle{ruled}
\newfloat{listing}{tb}{lst}{}
\floatname{listing}{Listing}
\title{Improving Ontology Requirements Engineering with OntoChat and Participatory Prompting}
\author{
    Yihang Zhao\textsuperscript{\rm 1}, 
    Bohui Zhang\textsuperscript{\rm 1}, 
    Xi Hu\textsuperscript{\rm 1}, 
    Shuyin Ouyang\textsuperscript{\rm 1}, 
    Jongmo Kim\textsuperscript{\rm 1},
    Nitisha Jain\textsuperscript{\rm 1},
    Jacopo de Berardinis\textsuperscript{\rm 2},
    Albert Meroño-Peñuela\textsuperscript{\rm 1},
    Elena Simperl\textsuperscript{\rm 1}
}
\affiliations{
    \textsuperscript{\rm 1}King's College London\\
    \textsuperscript{\rm 2}University of Liverpool\\
    \{yihang.zhao, bohui.zhang, jongmo.kim, xi.hu, shuyin.ouyang, jongmo.kim, nitisha.jain, albert.merono, elena.simperl\}@kcl.ac.uk, jacodb@liverpool.ac.uk
}

\usepackage{bibentry}
\usepackage{refcount}
\usepackage{natbib}

\begin{document}

\maketitle

\begin{abstract}
Past ontology requirements engineering (ORE) has primarily relied on manual methods, such as interviews and collaborative forums, to gather user requirements from domain experts, especially in large projects. Current OntoChat offers a framework for ORE that utilises large language models (LLMs) to streamline the process through four key functions: user story creation, competency question (CQ) extraction, CQ filtration and analysis, and ontology testing support. In OntoChat, users are expected to prompt the chatbot to generate user stories. However, preliminary evaluations revealed that they struggle to do this effectively. To address this issue, we experimented with a research method called participatory prompting, which involves researcher-mediated interactions to help users without deep knowledge of LLMs use the chatbot more effectively. This participatory prompting user study produces pre-defined prompt templates based on user queries, focusing on creating and refining personas, goals, scenarios, sample data, and data resources for user stories. These refined user stories will subsequently be converted into CQs.
\end{abstract}

\section{Introduction}

Ontology Requirements Engineering (ORE) plays a critical role in the design, evaluation, and reuse of ontologies, which are essential for structuring knowledge across various domains. Traditionally, ORE has relied heavily on manual activities such as interviews, collaborative forums, and discussion pages to gather user requirements from domain experts, particularly in large projects involving stakeholders from diverse backgrounds \cite{zhang2024ontochat}. These manual processes are often resource-intensive and complex, posing significant challenges in ensuring comprehensive and consistent requirement collection.

The central theme of ORE is competency questions (CQs). The Infer, DEsign, CreAte (IDEA) framework \cite{de2023polifonia} was an initial tool set that paved the way for language model-driven ontology engineering in large-scale projects. It automates the extraction, organisation, and refinement of CQs, detecting inconsistencies, and using visual and analytical tools for question management.

Ontological requirements are gathered from customers in the form of user stories, which are then translated into CQs \cite{de2023polifonia}. OntoChat supports this process by providing conversational agents that facilitate ORE, beginning with the creation of user stories. Users can create user stories by answering elicitation questions, and ontology engineers can extract competency questions (CQs) and test preliminary versions of ontologies. Despite these advancements, the researcher found that while this system was helpful, participants often needed more specific guidance and iterative refinement to produce satisfactory user stories.

This study addresses identified limitations by using participatory prompting user study protocols. Participants express ontology-related queries, which researchers refine using predefined prompting strategies \footnote{\label{fn:pre_identified_prompts}\url{https://github.com/King-s-Knowledge-Graph-Lab/OntoChat/blob/main/assets/user_study/Pre_identified_Prompts_for_Ontology_User_Story_Elicitation.md}}. We then used GPT-4o to generate responses, with participants providing feedback and posing additional queries. Researchers further refined these queries into prompts. This iterative cycle enhances query articulation and facilitates effective interaction with LLMs for user story creation. The results of this user study include pre-defined prompt templates based on user queries, which are available online \footnote{\label{fn:user_needs}\url{https://github.com/King-s-Knowledge-Graph-Lab/OntoChat/blob/main/assets/user_study/User_Needs_for_the_LLM_assisted_Task_Assisting_in_User_Story_Creation.md}}. These prompts can serve as elicitation questions for users or be personalised to seek assistance from OntoChat to improve the current ``Assisted Persona and Story Creation" workflow. In sum, this work contributed by applying the participatory prompting method to OntoChat to enhance the user experience. Using this approach, we developed pre-defined prompt templates tailored to user queries for creating personas, goals, scenarios, example data, and data resources within the ontology user story creation workflow.

\section{Related Work}

Various ORE methodologies, such as NeOn \cite{suarez2011neon}, SAMOD \cite{peroni2017simplified}, and eXtreme Design (XD) \cite{presutti2009extreme}, include tasks like eliciting user stories, defining purpose, scope, objectives, domain, coverage, and granularity, specifying formality levels, identifying end-users and intended uses, eliciting competency questions and expected answers, specifying non-functional requirements, gathering resources, validating competency questions, extracting glossaries of terms, grouping competency questions and assigning priorities, and reusing resources.

ORE typically begins with gathering requirements by identifying the domain, relevant content, stakeholders, existing systems, and data sources. A user-friendly way to express requirements is through ontology user stories, which use techniques from User Experience (UX) design \cite{de2023polifonia, presutti2009extreme, zhao2024userexperiencedatasetsearch}. An ontology user story includes a persona, described with details like name, age, occupation, skills, and interests to capture the user's perspective. The goal section outlines what the persona aims to achieve with the ontology, using up to five keywords to summarise main objectives. The scenario describes the current methods the persona uses, highlighting gaps and improvements the ontology can provide. Example data illustrates the types of information the persona will interact with, such as text, numbers, or images, with specific examples like book titles for a librarian or song titles for a music enthusiast. Data resources, such as websites, books, or tools that could assist the persona, may also be listed.

Recently, LLMs have been employed in knowledge engineering for tasks like representing domain knowledge, generating examples of classes and relations, providing explanations and recommendations on ontologies \cite{meyer2023llm}, ontology matching and alignment \cite{qiang2023agent, he2023exploring, hertling2023olala}, ontology refinement \cite{zhao2024using}, and ontology construction and learning from text \cite{funk2023towards}. In requirement elicitation, new methods have been developed for retrofitting and extracting competency questions (CQs) from ontologies and knowledge graphs \cite{ciroku2024revont}, automatically translating natural language domain descriptions into ontologies \cite{fathallahneon}, and using conversational agents to steer the creation of user stories \cite{zhang2024ontochat}.

To provide a detailed introduction to OntoChat user story creation \cite{zhang2024ontochat}, it facilitates a collaborative approach through a structured four-step process involving both back-end operations and user engagement. In Step 1 (back-end), the LLM is primed to act as a knowledge elicitor, designed to gather detailed information about each user story component (persona, goals, scenarios, and data examples). In Step 2 (user involvement), users are guided through specific questions to collect insights; for example, OntoChat asks, ``What are the name, occupations, skills, and interests of the user?" to define the persona. If responses are incomplete, the LLM prompts for additional details until all necessary information is gathered. In Step 3 (back-end), the LLM uses a one-shot learning approach to create an initial user story draft by following the provided example structure. Finally, in Step 4 (user involvement), users review the initial draft and provide feedback for iterative refinement, making corrections, adding details, or removing irrelevant information until the user story is finalised.

While these works show significant potential for LLM-driven knowledge engineering, they also acknowledge challenges such as generating inaccurate information (hallucination), the high costs associated with fine-tuning, and insufficient non-linguistic reasoning skills.

\section{Data and Methods}

To improve the user experience and develop pre-defined prompt templates tailored to user queries for the current OntoChat user story creation workflow, we conducted a three-step participatory prompting process \cite{sarkar2023participatory}: \textbf{(1) Story Description} - Participants wrote stories based on their ontological needs; \textbf{(2) Prompt Development} - Participants posed queries to request LLM support in writing user stories. Researchers refined these queries using pre-identified prompting strategies \footnotemark[\getrefnumber{fn:pre_identified_prompts}], then asked the LLM for answers, with participants evaluating the LLM-generated responses and providing feedback for further refinement; \textbf{(3) Story Evaluation} - Participants assessed the stories for usefulness, clarity, and inspiration, resulting in more detailed and practical user stories. Ethical clearance was obtained from the Department of Informatics at King's College London with reference number MRSP-23/24-41128, ensuring that no personal data was collected throughout the experiment. The complete user study script are available online \footnote{\url{https://github.com/King-s-Knowledge-Graph-Lab/OntoChat/blob/main/assets/user_study/Study_Script_for_User_Story_Writing.md}}. Below, we provide a detailed explanation of how each of these steps works.

\subsection{Participants}

This user study involved ten participants \footnote{\url{https://github.com/King-s-Knowledge-Graph-Lab/OntoChat/blob/main/assets/user_study/Demographic_Information_of_Participants.md}}, including MuseIT practitioners aiming to build an ontology for multi-sensory representation in cultural heritage and PhD researchers whose research focuses on knowledge engineering. They have expertise in areas such as multisensory interaction, cultural heritage, mixed reality, multimodal representation learning, machine learning, and responsible AI. Their familiarity with KGs, LLMs, ORE, and software RE varies, with many well-versed in KGs and LLMs. For example, PID7, a software engineer specialising in multisensory interaction, described their approach to creating user stories for integrating tactile data in multisensory museum exhibits.

\subsection{Step 1: Story Description}

Participants were first instructed on how to write ontology user stories and what constitutes a good user story based on the standard user story shown in Figure \ref{fig:user_story} and the provided instructions \footnote{\url{https://github.com/King-s-Knowledge-Graph-Lab/OntoChat/blob/main/assets/user_study/Instructions_on_How_to_Write_Ontology_User_Stories.md}}. For example, the instruction for writing the persona section: \textit{``When writing a Persona, imagine a character who represents the typical user of your ontology. This character is called a persona. Describe this persona with details like their name, age, job, and their skills or interests. This helps you consider how your ontology can be useful for people like them."}. Participants were tasked with writing user stories leveraging their domain-specific knowledge. For instance, PID7, a software engineer in multisensory interaction, uses the MuseIT ontology to connect cultural artifacts with visual, auditory, and tactile elements for a museum exhibit. Previously, this process was manual and time-consuming. This ontology allows them to efficiently map sensory inputs, ensuring the exhibit is cohesive, accessible, and enhances the visitor experience.

\subsection{Step 2: Prompt Development}

This process involved iterative ``turns" where participants interacted with a researcher and GPT-4o. Each turn began with the participant posing a query related to their ontology project user story, such as requesting help in generating or refining a part of the story. For instance, PID7 sought help in identifying a scenario where an ontology could enhance the design of multisensory museum exhibits, particularly in integrating tactile elements. The researcher then refined this query using a pre-identified prompting strategy \footnotemark[\getrefnumber{fn:pre_identified_prompts}], such as: \textit{``Describe the current ways [the persona] performs [actions], and ensure they align with the persona’s [occupation] and [skills]."} The placeholders were replaced with PID7’s specific details: ``software engineer" for [occupation], ``expert in Visual Arts and Interactive Media" for [skills], and ``enhancing the design of multisensory museum exhibits, particularly in integrating tactile elements" for [actions].

Participants provided feedback on the LLM's responses and requested further refinements or new queries. Initially, GPT-4o produced basic user stories with clear personas, goals, and scenarios but lacked detailed information. For instance, when participant PID7 asked for a story about integrating tactile data in multisensory exhibits, GPT-4o's responses were too general, only mentioning sensory integration without specifying details. As the sessions progressed, PID7 requested more specifics on data sources, haptic feedback models, and integration techniques. GPT-4o improved by incorporating details like tactile sensor data, haptic device compatibility, and sensory mapping methods, though further refinement was needed for practical applications. Participants sometimes requested more realistic elements, such as managing data interoperability and practical challenges. They also emphasised the query for interdisciplinary insights, reliable sources, and ethical considerations relevant to their domains. By addressing these queries, this study identified key elements for creating detailed and practical user stories for ontology requirements engineering using an AI assistant.

Throughout the participatory prompting sessions, researchers created specific user prompts to assist participants in critically evaluating the LLM's output and refining their queries for improved responses. These prompts encouraged participants to give feedback on the LLM's responses. For instance, a prompt might ask, ``What would you change in your query to make this more useful? Would you ask this a different way?" This iterative method ensured that the LLM's outputs became progressively more relevant and detailed, ultimately leading to the development of more practical and comprehensive user stories.

\subsection{Step 3: Story Evaluation}

Post-activity interviews were conducted to gather participants' satisfaction with the current user stories. Each participant rated the stories generated with GPT-4o based on their relevance, clarity, and usefulness using a Likert scale from 1 to 5, where 1 indicated strong dissatisfaction and 5 indicated strong satisfaction. All finalised prompts in our repository generated satisfactory results based on metrics of relevance, clarity, and usefulness, with each metric scoring 4 or higher, indicating high user satisfaction. Participants also identified any additional queries needed to improve the stories. If further queries were identified, the researcher would immediately try those queries and gather feedback again. This feedback loop helped refine the prompts and improve the overall quality of the user stories generated by OntoChat.

\subsection{Data Analysis}

After this user study, ten videos were manually transcribed, and the data was organised into five primary sections followed by a user story template, including ``Persona," ``Goal," ``Scenario," ``Example Data," and ``Data Resource." The coding process combined top-down and bottom-up approaches, as outlined by \cite{gu2014code}. Codes were designed to be overlapping to capture the multifaceted nature of user queries. To confirm reliability, the initial coding was reviewed at a later stage, and any deviations were corrected through iterative revisions.

\section{Results}
\begin{figure}[ht]
    \centering
    \includegraphics[width=1\columnwidth]{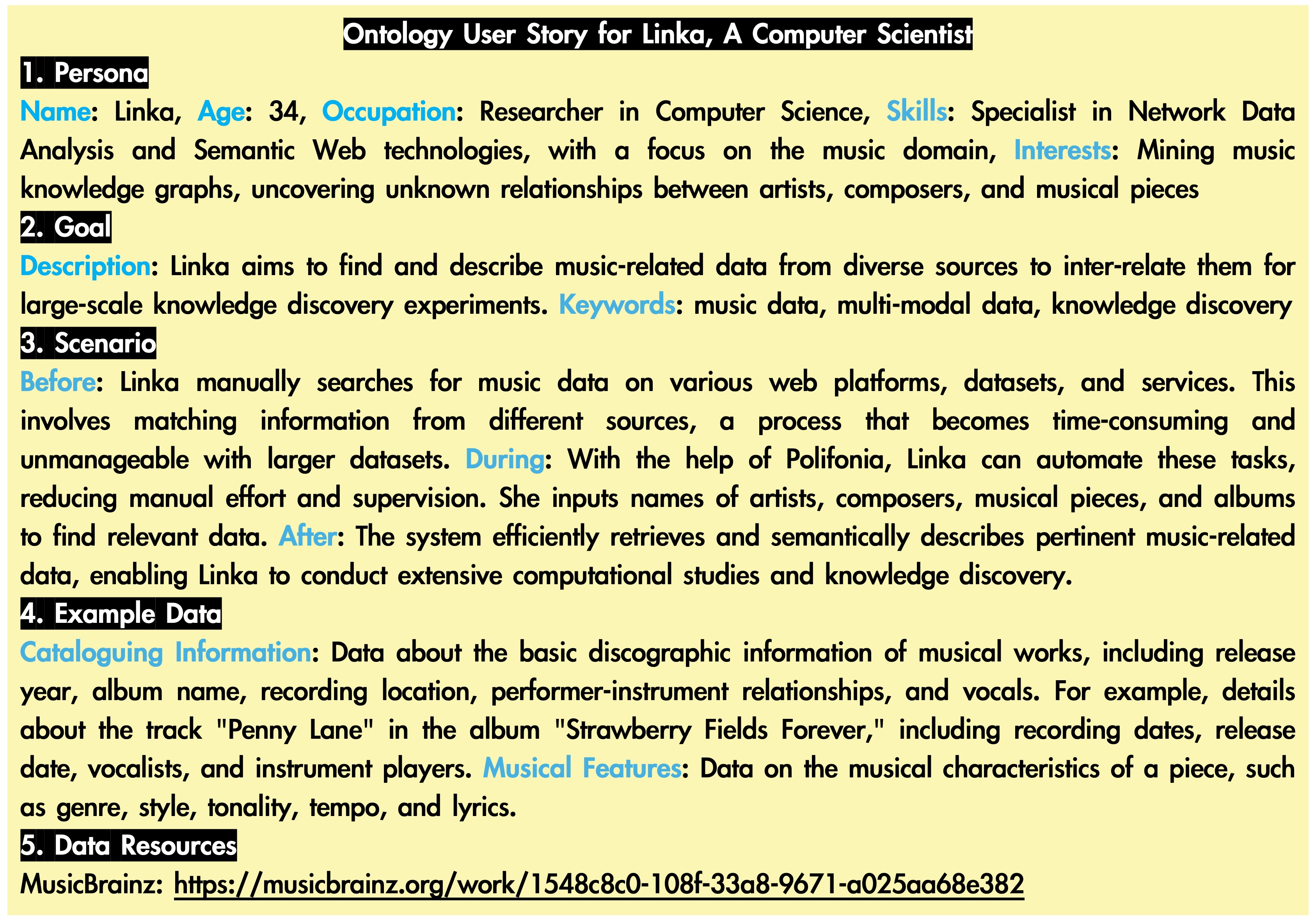}
    \caption{An example of a standard ontology user story}
    \label{fig:user_story}
\end{figure}

The full list of user queries and corresponding prompts is available online \footnotemark[\getrefnumber{fn:user_needs}]. These prompts were developed for each user query through iterative refinement with researchers and participants, who found that they effectively guide the system in generating an ontology user story of satisfactory quality, as shown in the example in Figure \ref{fig:user_story}. The results section summaries user needs for each part of an ontology user story, providing an example user query and its related prompt for each part.

\subsection{Persona}

For the persona section, users ask for LLM help to refine traits like occupation, skills, and interests, making sure they are detailed, interconnected, and aligned with the project. This helps users create more accurate goals and scenarios. For example, \textbf{user query}: \textit{``Detail one persona trait."} \textbf{prompt}: \textit{``Could you improve the persona by detailing [this specific trait] needed for writing an ontology user story? Ensure it connects with other traits and fits the project’s needs."}

\subsection{Goal}

For the goal section, users ask for LLM help to write a clear goal description with relevant keywords and the actions the persona needs to take, ensuring they align with the persona traits. To enhance the goal, they ask the LLM to categorise it into short-term and long-term objectives, provide insights on methodologies, tools, and interdisciplinary factors. For example, \textbf{user query}: \textit{``Describe the goal."} \textbf{prompt}: \textit{``Could you write a clear goal description with the sequence of [actions] for this persona? Ensure the goals align with the persona's [interests] and the [actions] fit the persona’s [occupation] and [skills]."}

\subsection{Scenario}

For the scenario section, users sought LLM assistance with tasks such as including the current methods, challenges in achieving goals, new methods for improvement, and expected outcomes. They also aimed to enhance the scenario section by asking the LLM to outline the sequence of actions, identify key players, and describe their interactions. For example, \textbf{user query}: \textit{``Narrative story for scenario."} \textbf{prompt}: \textit{``Could you narrate a detailed story about the scenario? Include the [current ways] the persona performs [actions], the [challenges] encountered, and how the ontology introduces [new ways] to address these challenges. Describe the [outcomes] after using the ontology. Review the scenario to ensure the [current ways] align with the [actions] in the goal section and the [outcomes] relate to the [user goal]. Ensure that the [current ways] and [new ways] match the persona’s [occupation] and [skills], and that the [outcomes] align with the persona’s [interests]."}

\subsection{Example Data}

For the example data section, users requested LLM assistance with tasks such as providing a general description of the data category, explaining relationships within the category, and identifying historical details like how the category has evolved over time. For example, \textbf{user query}: \textit{``Detailed description of one category."} \textbf{prompt}: \textit{``Could you provide a detailed description of the characteristics of [this specific category item]? Explain how it relates to the persona's traits, goals, or scenarios."}

\subsection{Data Resource}

For the data resource section, users requested LLM assistance with tasks such as suggesting primary data sources, specifying data formats, and detailing the types of data included with a given source link. They also needed the LLM to provide details on the dataset metadata, access methods, and any restrictions associated with a given source. For example, \textbf{user query}: \textit{``Check metadata quality of one data source."} \textbf{prompt}: \textit{``Could you provide a detailed description of the metadata information available in [this source]? Please include quotes from the cited source to support your explanation."}

\section{Discussion}

Even with participatory prompting, generating effective prompts was a lengthy and iterative process. This involved multiple cycles of participants asking queries to write user stories, researchers converting these queries into prompts, and participants providing feedback on the LLM's responses. Knowing that every new query would be formally converted into a prompt, participants carefully considered each query before presenting it, resulting in some potential ideas being overlooked and untested. To address this, we encouraged and guided participants to actively provide feedback on the LLM's responses.

\section{Future Works}

We propose enhancing OntoChat's user story refinement workflow by incorporating pre-defined prompt templates and an LLM to suggest targeted prompts. This will guide users in improving specific aspects of their stories. A systematic A/B test will compare OntoChat v1.0 (current version) with v2.0, featuring these new functions.

\section{Conclusion}

In this study, we used participatory prompting protocols to enhance the user experience and create pre-defined prompt templates for ontology user story creation with LLMs. This method addresses challenges non-experts face with effective prompting. We also proposed new workflows to improve OntoChat's user story refinement by providing suggested prompts, which users can respond to or personalise. This structured approach helps users identify improvements and articulate queries effectively, enhancing user story quality and ontology engineering practices.

\section{Acknowledgements}
This work was partially supported by the MuseIT project, co-funded by the European Union through the Horizon Europe Framework Program (Grant Agreement No. 101061441). The views and opinions expressed in this work are those of the authors and do not necessarily reflect the views of the European Union or the European Research Executive Agency.

\bibliography{aaai24}

\end{document}